# Charge Transport Through DNA with Energy-Dependent Decoherence

Hashem Mohammad[1*] and M. P. Anantram[2]

[1] Department of Electrical Engineering, Kuwait University, P.O. Box 5969, Safat 13060, Kuwait

[2] Department of Electrical and Computer Engineering, University of Washington, Seattle, WA 98195

*Corresponding author, email: hashem.mohammad@ku.edu.kw

**Abstract**

Modeling charge transport in DNA is essential to understand and control the electrical properties and develop DNA-based nanoelectronics. DNA is a fluctuating molecule that exists in a solvent environment, which makes the electron susceptible to decoherence. While knowledge of the Hamiltonian responsible for decoherence will provide a microscopic description, the interactions are complex and methods to calculate decoherence are unclear. One prominent phenomenological model to include decoherence is through fictitious probes that depend on spatially variant scattering rates. However, the built-in energy-independence of the decoherence (*E-itbndep*) model overestimates the transmission in the bandgap and washes out distinct features inside the valence or conduction bands. In this study, we introduce a related model where the decoherence rate is energy-dependent (*E-dep*). This decoherence rate is maximum at energy levels and decays away from these energies. Our results show that the *E-dep* model allows for exponential transmission decay with the DNA length and maintains features within the bands' transmission spectra. We further demonstrate that we can obtain DNA conductance values within the experimental range. The new model can help study and design nanoelectronics devices that utilize weakly-coupled molecular structures such as DNA.

# 1. Introduction & Background

Understanding and controlling the electrical conductivity of nucleic acids (DNA) based structures has gained interest in the past decade due to advances in electrical methods for disease detection in biology and the development of DNA origami. Measuring DNA conductance for sensing biological processes [1–4], developing new sequencing techniques[5–7], and future molecular device applications [8–17] have led to an interest in its electrical properties. DNA origami exploits the self-assembly property of DNA to create complex three-dimensional architectures [18]. This technique helps build nanoscale structures bottom-up instead of the top-down approach currently used in nanoelectronics. Further, the ability to build heterostructures [19] and dope [14–17] DNA makes it a potential candidate for nanoelectronics beyond lithography. Therefore, understanding how charge transports through nucleic acids can help engineer a new class of biosensors and nanoelectronics[3,13,20–22]. From a fundamental viewpoint, the precise mechanism of charge transport in DNA is not fully understood.

In probing the electrical transport, we consider a system that consists of two metal contacts with DNA in between them (Figure 1a). It has been shown previously that coherent transport calculations yield conductance values that are orders of magnitude smaller than experimental values [23]. The calculated conductance becomes comparable to experiments by including decoherence to account for the realistic effects mentioned above. Decoherence has been modeled at various levels of theory. Gutiérrez et al. used a dissipative phonon-bath technique where the base is represented by a single energy level, and this energy level is coupled to a backbone which is also represented by a single energy level. Only the backbone's energy level is coupled to the phonon bath [24,25], Kubar et al. model decoherence by solving time-dependent equations that account for the variation in the energy levels of DNA bases with time [26], while Karasch et al. model decoherence due to including vibronic dephasing [27]. These methods to include coupling to a phonon/vibronic bath include only one to three energy levels on the DNA bases. Although these models can qualitatively describe the small hopping parameter between DNA bases, using the full Hamiltonian is important to explain the impact of the backbone, methylation of bases [3], single nucleotide mismatches [4], and intercalation [21]. These theoretical methods that probe realistic DNA strands with many hundreds of atoms by accounting for the (i) backbone, (ii) sequence, (iii) transport of electrons between the complementary strands, and (iv) environment consisting of a solvent are essential to correctly model electron transport in DNA and toy models do not suffice[28,29].

In our approach here, the Hamiltonian accounts for all coherent scattering events involving multiple energy levels and the hopping energy between energy levels at many bases. It is not computationally feasible to treat decoherence in such a model using non perturbative methods; perturbative methods fail in weakly coupled systems such as DNA. Our model to include decoherence includes the effect of elastic collisions by using *decoherence probes*. While our empirical model doesn't include individual phonon interactions, a self-energy term accounts for decoherence with an environment. The decoherence probe method originated in the late 1980s. It was originally proposed by H. L. Engquist and P. W. Anderson [30] and later used efficiently by Büttiker to model nanostructures [31] as a way of including environmental interactions using multiple added contacts. This method was then utilized extensively in nanoscale transport simulators. While the decoherence probe method is not without drawbacks, it does an exceptionally good job of enabling current continuity with self-energies in the non-equilibrium context, where the values of these self-energies can be generated from experimental results or theoretical models. Further, the decoherence probes adopted in the cited publications use an energy independent decoherence rate. In this work we go beyond by making the decoherence energy dependent in this phenomenological model. These decoherence probes are physically modeled as self-energies in a Green's function formalism, which is used to calculate the conductance [23,32,33]. Central to this model are spatially dependent decoherence parameters that represent the average scattering time in which an electron loses its phase information. The model is adept at including the effect of density of states (DOS) broadening. In the coherent limit, the DOS peaks along the length of the DNA have almost no energetic overlap, making transport difficult. The decoherence probes broadens the DOS peaks significantly and hence increases the transmission probability across the DNA length [23]. The decoherence probe model has qualitatively explained experiments and described different charge transport mechanisms through DNA such as direct tunneling, sequential hopping, and intermediate coherent-decoherent regime [4,23,33–36].

A major drawback of prior work is that the decoherence rates are *energy-independent*, an unphysical assumption. While this assumption simplifies the problem considerably, allowing one to neglect the real part of the self-energy because it is precisely zero, it yields some qualitatively incorrect results. First, the transmission inside the bandgap is found to be unphysically large. As a result, there is only a weak dependence of transmission as a function of length in the bandgap. Second, interesting features in the transmission are washed out within the conduction and valence bands. When the *energy-independent* decoherence model is used in a crystalline semiconductor, to make the transmission small in the bandgap, the decoherence rate is set to zero (or a small value). This treatment is unphysical because it does not conserve the norm of the DOS integrated over all energies.

To summarize, in modeling transport through DNA, we must carefully consider the impact of the decoherence model on both the transmission and the integral of the DOS. Notably, the model should allow for exponential decay of transmission with the length within the bandgap and maintain features in the transmission spectra within the bands. In this paper, we modify the phenomenological decoherence model to make the decoherence rate energy-dependent. The role of the real part of the self-energy on both the transmission and the integral of DOS are also discussed.

The paper is organized as follows. Section 2 discusses the methodology and the model. We model the DNA using density functional theory (DFT) with the polarizable continuum model (PCM) to account for the solvent dielectric constant. We then calculate the transmission and conductance using the Green's function method with decoherence probes to account for the decoherence. Following this, we compare the *energy-dependent* and *energy-independent* decoherence models with a *model Hamiltonian* for the DNA in section 3. We then discuss the decoherence model and the role of the real part of the self-energy in sections 4 and 5. Finally, we discuss the transport results with the full Hamiltonian of the double stranded DNA and compare the results of our model to experiments in section 6.

## 2. Methods

We study the double-strand B-DNA sequence 3'-$C_3G[CG]_lC_3$-5', with $l$ = 1-4 in a contact-DNA-contact setup as shown in Figure 1a. We model the DNA using DFT to generate the system Hamiltonian and calculate transmission and conductance using Green's function method with decoherence probes to account for the decoherence. For the charge transport calculations, we assume the contacts to be at the cytosine nucleotides of both 3'- and 5'- ends.

We obtain the atomic structure of the B-DNA strands using the nucleic acid builder in Amber [37]. Using the coordinates obtained, we perform single-point DFT calculations using the B3LYP functional and 6-31G(d,p) basis set [38]. We include the effect of water solvent around the DNA *via* the polarizable continuum model (PCM) using the integral equation formalism (IEFPCM). Several studies find that the B3LYP with the 6-31G(d,p) basis set yields the same trend for ionization potentials (IP) as experiments, with an offset in values of about 300 meV [39]. Other methods such as MP2 and CCSD [40,41] may better match experimental IP results, but they are more computationally expensive. In addition, Felix and Voityuk [42] benchmarked the performance of DFT hybrid and nonhybrid functionals for hole transfer parameters in DNA. They used the smallest basis set for atomic orbital polarizability (the 6-31G(d) basis set). They found that hybrid functionals such as B3LYP have better accuracy than nonhybrid functionals. We use the 6-31G(d,p) in our calculations, which further adds the p-type polarization to the hydrogen atoms to better describe the hydrogen bonds present in the DNA base-pairs. Moreover, ab-initio studies have shown that the polarizability of the solvent that surrounds the molecule is essential to obtaining the correct energy levels and IP [41,43–45]. The PCM model adequately accounts for the dielectric constant of the solvent and the associated screening of charge. For example, Slavicek et al. [41] have shown that modeling water through PCM yields DNA IP results that directly correlate with photoelectron spectroscopy measurements without including explicit water molecules.

The systems we model contain more than 1000 atoms, and our charge transport model uses the full Hamiltonian to account for interactions between orbitals centered on atoms at different bases. Therefore, our choice of B3LYP/6-31G(d,p) as the calculation method is to balance between accuracy and computational efficiency given our computational resources. We have also included comparison with B3LYP/3-21G and B3LYP/cc-pVDZ in Appendix A.

Our calculations do not include the counterions surrounding the DNA because we are using a high dielectric solvent (the water dielectric constant is 78.4). The high dielectric constant screens the counterions surrounding the DNA and limits their role to only adding unoccupied energy levels deep in the LUMO regions [46]. Thus, the DNA transmission near HOMO or LUMO in the bandgap is not affected by the counterions. Since we omitted the counterions that neutralize the DNA backbone, we set the total charge equal to the number of phosphate groups in the DNA strands. The converged calculations result in the Fock ($F$) and overlap ($S$) matrices, which are then used as input in the charge transport calculations. In the following, we discuss the charge transport calculation method.

We use a Löwdin transformation to obtain the Hamiltonian ($H_0$) in an orthogonal basis set

$$H_0 = S^{-\frac{1}{2}} F S^{-\frac{1}{2}} \tag{1}$$

The Hamiltonian of the DNA (bases and backbone) can be expressed as

$$H_0 = H_0^D + H_0^{OD} \tag{2}$$

$$H_0^D = \sum_{\substack{\alpha=1 \to b_i \\ i=1 \to N_{DNA}}} \epsilon_{\alpha,i} \, c_{\alpha,i}^\dagger \, c_{\alpha,i} \tag{3}$$

$$H_0^{OD} = \sum_{\substack{\alpha=1 \to b_i \\ \alpha'=1 \to b_j \\ i,j=1 \to N_{DNA}}} t_{\alpha,i \to \alpha',j} \, (c_{\alpha,i}^\dagger \, c_{\alpha',j} + H.c) \tag{4}$$

Here, $\epsilon_{\alpha,i}$ stands for the on-site energy $\alpha$ of the $i^{th}$ atom, $c^\dagger$ and $c$ are the creation and annihilation operators, respectively. $t_{\alpha,i \to \alpha',j}$ is the interaction between orbital $\alpha$ at atom $i$ and orbital $\alpha'$ at atom $j$, and $H.c$ is the Hermitian conjugate. $b_i$ is the number of basis functions representing atom $i$, $N_{DNA}$ is the total number of atoms in the DNA strand.

In terms of matrices, one can think of $H_0^D$ ($H_0^{OD}$) as a diagonal (off-diagonal) matrix. The diagonal elements of $H_0$ represent the energy levels at each atomic orbital, and the off-diagonal elements correspond to the hopping parameters between them. The Hamiltonian $H_0$ is a square matrix with a dimension equal to the total number of basis functions in the system ($\sum_{i=1}^{N_{DNA}} b_i$).

We chose to partition the DNA into nucleotides (backbone + base) as shown in Figure 1a, where nucleotide $m$ corresponds to the partition labeled $m$. The $m$-th diagonal block of the Hamiltonian $H_0$ corresponds to the sub-Hamiltonian of nucleotide $m$.

Let $H_{0m}$ represent the diagonal blocks of $H_0$ corresponding to nucleotide $m$, and $u_m$ is a diagonal sub-matrix consisting of the eigenvectors of $H_{0m}$. We define the matrix $U$ to be

$$U = \begin{bmatrix} u_1 & 0 & & \cdots & & 0 \\ 0 & u_2 & & & & \\ \vdots & & \ddots & & & \vdots \\ & & & u_m & & \\ & & & & \ddots & 0 \\ 0 & & \cdots & & 0 & u_N \end{bmatrix} \qquad (5)$$

where $N$ is the total number of nucleotides in the DNA and the dimension of $u_m$ is $O_m$, and,

$$O_m = \sum_{i=1}^{N_m} b_i \qquad (6)$$

is equal to the total number of basis functions ($O_m$) used to represent nucleotide $m$. $N_m$ is the total number of atoms in nucleotide $m$. Then, the unitary transformation,

$$H = U^\dagger H_0 U \qquad (7)$$

diagonalizes the diagonal blocks corresponding to each nucleotide $m$. The diagonal blocks of $H$ are now diagonal matrices (see the matrix in Figure 1b). The diagonal elements of the diagonal blocks represent the eigenvalues of that nucleotide. The off-diagonal blocks of $H$ represent the hopping parameters between the molecular orbitals of the nucleotides. The size of each diagonal block corresponding to nucleotide $m$ is equal to $O_m$.

The redefined Hamiltonian $H$ in second quantized notation is given by:

$$H = H^D + H^{OD} \tag{8}$$

$$H^D = \sum_{\substack{k=1 \to O_m \\ m=1 \to N}} \epsilon_{k,m}\, c^\dagger_{k,m}\, c_{k,m} \tag{9}$$

$$H^{OD} = \sum_{\substack{k=1 \to O_m \\ k'=1 \to O_{m'} \\ m \neq m' = 1 \to N}} t_{k,m \to k',m'}\, (c^\dagger_{k,m} c_{k',m'} + H.c) \tag{10}$$

where $\epsilon_{k,m}$ is the $k^{\text{th}}$ on-site energy (molecular orbital) of the $m^{\text{th}}$ nucleotide, and $t_{k,m \to k',m'}$ is the interaction between the $k^{\text{th}}$ molecular orbital of the $m^{\text{th}}$ nucleotide with the $k'^{\text{th}}$ molecular orbital of the $m'^{\text{th}}$ nucleotide.

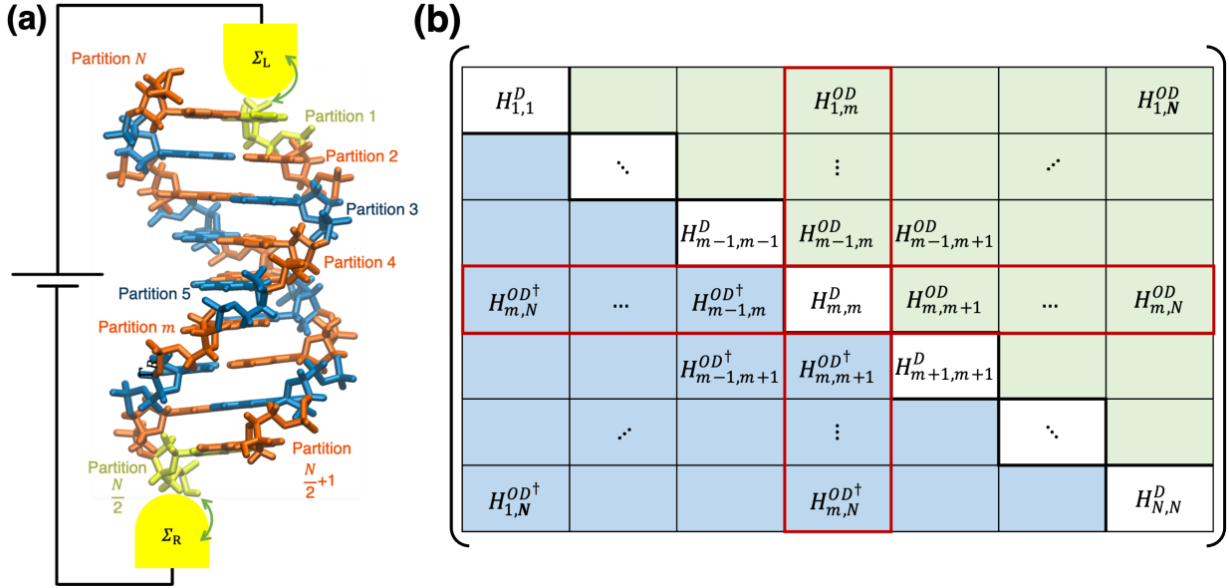

Figure 1 (a) Schematic of the charge transport calculations setup. The ds-DNA is connected to two electrodes through the 3'- and 5'- ends (nucleotides colored in yellow). The two partitions colored in yellow define the contacts self-energy locations at the 3'- and 5'-end nucleotides. The other two alternating colors of the DNA nucleotides resemble the partitioning scheme considered in the decoherent transport model. (b) The Hamiltonian $H$ after the partitioning into nucleotides per partition. The diagonal blocks contain the on-site potentials of each partition, and the off-diagonal blocks contain the hopping parameters between the partitions (in green and blue, where blue is the Hermitian conjugate of the green). The highlighted partition $m$ in red is to help guide the reader to its row and column. The green (or blue) portion of the $m^{\text{th}}$ row and column contains the hopping parameters between partition $m$ with all the other partitions in the system.

The Hamiltonian $H$ defined by equation 8 above is used to find the retarded Green's function ($G^r$) of the DNA by including both the self-energies due to decoherence and the contacts,

$$[E - (H + \Sigma_L + \Sigma_R + \Sigma_B)]G^r = I \quad (11)$$

where $E$ is the energy. $\Sigma_{L(R)}$ is the left (right) contact self-energy, which represents the coupling of the DNA to the left (right) contacts through which electrons enter and leave the DNA (Figure 1a). The self-energy for decoherence is represented by $\Sigma_B$. Both contact and decoherence self-energies are diagonal matrices, which means they are only coupled to $H^D$ (the molecular orbitals) in equation 9. We implement the self-energies in accordance with the partitioning scheme; each decohrence probe is connected to each nucleotide except for the two nucleotides attached to the two contacts (nucleotides 1 and N/2 in Figure 1a). We note that in the blocking scheme discussed above, there are spurious eigenvalues resulting from the dangling bond at the backbone between each neighboring nucleotide. We neglect these spurious energy levels and do not apply decoherence probes to them (see Appendix B for further details).

The matrix representation of the decoherence self-energy for the structure in Figure 1a is then

$$\Sigma_{B_{mm}}(k, k') = \Sigma_{k,m}\, \delta_{kk'} \quad \forall\, m \neq 1, N/2 \quad (12)$$

$\Sigma_{k,m}$ represents the decoherence self-energy associated with molecular orbital $k$ of nucleotide $m$. All other elements of $\Sigma_B$ not defined in equation (12) are zero. The blocks of $\Sigma_B$ associated with the two contact blocks do not have decoherence probes attached to them and hence are zero.

To model decoherence, we consider both the real and imaginary parts of the self-energy,

$$\Sigma_B(E) = Re[\Sigma_B(E)] + i \times Im[\Sigma_B(E)] \quad (13)$$

where $i$ is the imaginary unit $\sqrt{-1}$. The real part is a measure of an energy-dependent shift in the on-site potential (molecular orbital). The imaginary part of the self-energy is a measure of electron flow between the DNA and decoherence probe. Both parts are related to each other by the Kramers-Kronig relation [47],

$$Re[\Sigma_B(E)] = \frac{1}{\pi} P \int \frac{Im[\Sigma_B(E')]}{E' - E} dE' \tag{14}$$

where $P$ is the Cauchy principal value. We define the imaginary part at each diagonal element as

$$Im[\Sigma_B(E)] = -\frac{\Gamma_{k,m}(E)}{2} \delta_{kk'} \delta_{mm'} \tag{15}$$

where $\Gamma_{k,m}$ represents the coupling strength between the decoherence probe and molecular orbital $k$ at nucleotide $m$. We note that we omitted the indices of the left-hand side for clarity.

In this paper, we take the imaginary part of decoherence $Im[\Sigma_B(E)]$ to be *energy-dependent (E-dep)*. We then compare the results obtained with the often-used model that assumes $Im[\Sigma_B(E)]$ to be *energy-independent (E-indep)*. For the *E-dep* model, we define $\Gamma_k$ to exponentially decay with energy ($E$) at each molecular orbital as,

$$\Gamma_{k,m}(E) = \Gamma_B \times \exp\left[-\frac{|E - \epsilon_{k,m}|}{\lambda}\right] \tag{16}$$

where $\Gamma_B$ determines the maximum value of decoherence strength and $\lambda$ is an exponential decay parameter that dictates how quickly the decoherence decays away from an energy level. The net effect of these two parameters determines how the decoherence (and broadening) varies with energy.

The *energy-independent* decoherence model corresponds to having $\lambda = \infty$ in equation (16), which makes $\Gamma_{k,m}(E) = \Gamma_B$, a constant. Substituting this into (14) gives,

$$Re[\Sigma_B(E)] = \frac{\Gamma_B}{\pi} P \int \frac{1}{E' - E} dE' = 0 \tag{17}$$

That is, the real part of the self-energy is zero if the imaginary part is a constant independent of energy.

For the self-energy of the contacts, we use the wide-band limit (WBL), where the real part of the self-energy is zero and the imaginary part is a constant independent of energy. This approximation stands when the DOS is almost a constant throughout the metal, which is almost true for gold [48]. The contact self-energies within these approximations are

$$\Sigma_{L(R)} = -\frac{i\Gamma_{L(R)}}{2} \delta_{kk'} \delta_{mm'}, \quad m, m' = L(R), \text{ or } 1(N/2) \text{ as in Figure 1a} \tag{18}$$

After setting up the Green's function equation (11), the density of states at nucleotide $m$ and energy $E$ can be computed by extracting the corresponding diagonal elements of the retarded Green's function

$$DOS(m, E) = -\frac{Im[diag(G_m^r(E))]}{\pi} \tag{19}$$

We will now discuss some detail of the model for decoherence following references [49] and [50]. In the low-bias regime, the current at the $m^{th}$ probe is

$$I_m = \frac{2q}{h} \sum_{n=1}^{N} \int_{-\infty}^{+\infty} T_{mn}(E)[f_m(E) - f_n(E)]dE = \int_{-\infty}^{+\infty} J_m(E)dE \tag{20}$$

In linear response, this can be simplified to

$$I_m = \frac{2q}{h} \sum_{n=1}^{N} T_{mn}[\mu_m - \mu_n], \quad m: 1 \to N \tag{21}$$

$$T_{mn}(E) = \text{Tr}\big[(-2 \times Im[\Sigma_B(E)]_{m,m}) \times G_{m,n}^r \times (-2 \times Im[\Sigma_B(E)]_{n,n}) \times G_{n,m}^a\big] \tag{22}$$

where $q$ is the elementary charge, $h$ is the Planck's constant, $T_{mn}$ is the transmission between the $m^{th}$ and $n^{th}$ probes, and $G^a = (G^r)^\dagger$ is the advanced Green's function. $f_m(E) = \left[1 + \exp\left(\frac{E - E_{fm}}{kT}\right)\right]^{-1}$ is the Fermi distribution, and $J_m(E)$ is the current per unit energy at probe $m$. In equation (22), the terms in parentheses correspond to the coupling strength to the contact or the decoherence probes (as per equations (15) and (18)). $G_{m,n}^r$ and $G_{n,m}^a$ are $(b_m \times b_n)$ and $(b_n \times b_m)$ matrices, respectively. In this work, we calculate the transmission at room temperature, thus, $kT = 0.0259$ eV.

Because the decoherence probes are fictious probes used to model decoherence, there is no net current flowing through them. As a result, at each decoherence probe, $J_m(E) = 0$. This yields $N_B$ independent equations from which the following relation can be derived

$$\mu_m - \mu_L = \left(\sum_{m=1}^{N_B} W_{mn}^{-1} T_{nR}\right)(\mu_R - \mu_L), \quad n = 1 \to N_B \tag{23}$$

Here, $W_{mn}^{-1}$ is the inverse of $W_{mn} = (1 - R_{mm})\delta_{mn} - T_{mn}(1 - \delta_{mn})$, where $R_{mm}$ is the reflection probability at probe $m$, and is given by $R_{mm} = 1 - \sum_{m \neq n}^{N} T_{mn}$. The currents at the left ($I_L$) and right ($I_R$) obey the relation $I_L + I_R = 0$. This yields the equation for the current at the left contact as

$$I_L = \frac{2q}{h} T_{eff}(\mu_L - \mu_R) \qquad (24)$$

Comparing equations (21) to 24 yields the effective transmission term as

$$T_{eff} = T_{LR} + \sum_{m,n}^{N_B} T_{Lm} W_{mn}^{-1} T_{nR} \qquad (25)$$

where the first term is the coherent transmission from the left electrode to the right electrode. The second term is the decoherence contribution into the transmission via the decoherence probes. The low-bias conductance as a function of Fermi energy ($E_f$) is calculated as

$$G(E_f) = \frac{2q^2}{h} \int dE \, T_{eff}(E) \frac{\partial f(E - E_f)}{\partial E} \qquad (26)$$

## 3. Exponential Decay with Length

To demonstrate the importance of the *E-dep* model, we use a *model Hamiltonian* for the DNA where each block is a base to calculate the transmission as a function of length. The ds-DNA typically has a HOMO-LUMO gap of 3 to 4 eV and hopping parameter between nearest neighbor base-pairs of 10-100 meV[22,51,52]. We can thus consider the DNA as a weakly-coupled wide bandgap semiconductor. With this large energy gap, we expect the transmission near the middle of the gap (i.e., the midgap) to drop exponentially with an increase in strand length. In this model Hamiltonian, each base has two energy levels to represent HOMO and LUMO (every $H^D$ and $H^{OD}$ in Figure 1b is now a 2x2 matrix). We create the model Hamiltonian by the following steps: (i) taking smaller subsequences of the larger studied sequences, (ii) generate their Hamiltonians using DFT, then (iii) extract the hopping parameters between the nearest neighboring bases with respect to the HOMO and LUMO. We first run DFT calculations for the following 3 base-pair strands: 3'-CCC-5', 3'-CCG-5', 3'-CGC-5', 3'-GCG-5', and 3'-GCC-5'. From the full DFT Hamiltonian we find the simplified Hamiltonian. This enables us to extract the necessary hopping parameters between the nearest neighboring blocks with respect to the HOMO and LUMO. The sequences generated using the model Hamiltonian are: 3'-$C_3$-5', 3'-$C_6$-5', and 3'-$C_3G[CG]_lC_3$-5', with $l$ = 1-4. These sequences are 3, 6, 9, 11, 13, and 15 base-pairs long, respectively.

Using the 3 base-pair strands, we generate the Hamiltonian of the studied sequences by extracting the parameters for one base-pair at a time, moving from 3'-end to the 5'-end. To demonstrate, we look at the first four bases highlighted in bold in 3'- **CCCG**CGCCC-5' ($l$ = 1 case) as an example. The parameters for the left-most 3'-end cytosine in bold are extracted from the 3'-CCC-5' sequence, along with its complementary guanine base, as shown in Figure 2a. The extracted data are the HOMO and LUMO on-site energies of the left-most cytosine and guanine, and their hopping parameters to the HOMO and LUMO of the nearest neighbor cytosine and guanine. The second cytosine in bold is surrounded by two cytosine bases. Therefore, the extracted parameters are the HOMO and LUMO on-site energies of the middle cytosine and its complementary guanine in 3'-CCC-5' with respect to their left-connected CG base-pair (Figure 2b). As for the third cytosine highlighted in bold, the parameters are now extracted from the 3'-CCG-5' sequence (Figure 2c). The parameters for the guanine highlighted in bold are extracted from 3'-CGC-5'(Figure 2d). We continue with this operation for all the bases in the studied sequences and end up with the nearest neighbor model Hamiltonian per sequence. To run the transport calculations, for simplicity, we set the left (right) contacts coupling $\Gamma_{L(R)}$ = 600 meV at the 3'- and 5'- end terminal blocks of this system (as in Figure 1a).

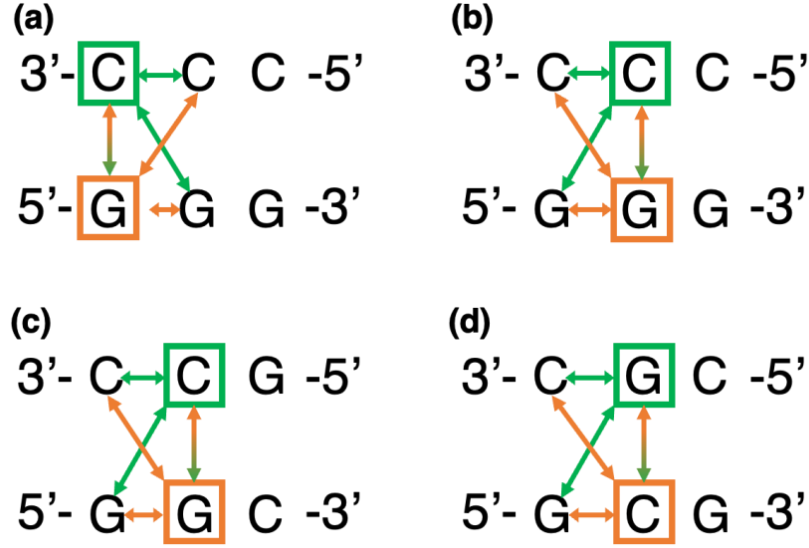

Figure 2 Examples of extracting the parameters from the 3 base-pair strands to generate the longer studied strands. **(a)** The HOMO and LUMO of the 3'-end cytosine (highlighted in green) and its complementary guanine (highlighted in orange) are extracted from the Hamiltonian. The hopping parameters to their nearest neighbor cytosine and guanine are also extracted (arrows). **(b)** The HOMO and LUMO of the middle cytosine (highlighted in green) and its complementary guanine (highlighted in orange) are extracted from the Hamiltonian. The hopping parameters to their left-connected cytosine and guanine are also extracted (arrows). **(c,d)** Using a different sequence combination to extract the parameters for a cytosine and guanine, respectively.

We calculate the transmission in the coherent limit and using *E-indep* and *E-dep* decoherent models. For both models, we set the decoherence rate to $\Gamma_B = 100$ meV. In the *E-dep* decoherent model, we vary the energy decay factor $\lambda = [10, 50, 100, 500]$ meV. We plot the results for a DNA strand with length equal to 9 base-pairs (bps) in Figure 3a, showing that the coherent transmission decays as we enter the bandgap region. Similarly, the *E-dep* model has an exponentially decaying tail into the bandgap (from HOMO or LUMO edges). We also note that increasing $\lambda$ can cause the *E-dep* model to deviate from the coherent limit deep in the bandgap. For low values of $\lambda$ (10 meV), the E-dep model deep in the bandgap follows the coherent limit. With moderate values of $\lambda$ (50 and 100 meV), the E-dep model starts to vary in the bandgap but follows the coherent limit at midgap. However, for large values of $\lambda$ (500 meV), the *E-dep* model yields similar values to the *E-indep* model, which is unphysical. This behavior is because increasing $\lambda$ makes the decoherence rate larger off-resonance (away from $\epsilon_r$) as seen in the exponential-decaying factor of equation 16. As for the *E-indep* model, it does not maintain the decay deep in the bandgap. Initially, the transmission drops as we enter the bandgap but then saturates -within 200 meV from HOMO or LUMO- at a relatively high value (T>$10^{-7}$).

Next, we plot the transmission extracted at the midgap as a function of length in Figure 3b. At λ = 10 and 50 meV, the *E-dep* model and coherent transport have comparable exponential decay in transmission with increasing length. Increasing λ to 100 meV lowers the decay at longer strands. Similar to the impact of large λ values (500 meV), the transmission in the *E-indep* model only decreases by a small amount for length ≥ 6 bps. Even at low decoherence rates such as 10 meV, the transmission of the *E-indep* model is unrealistically high at energies deep in the bandgap.

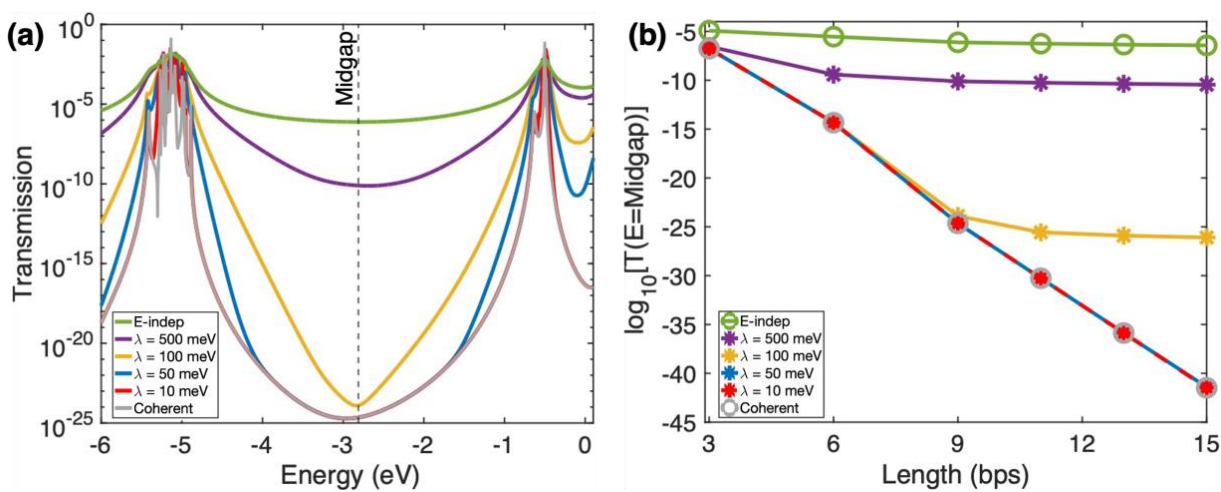

Figure 3 **(a)** Transmission plot for the DNA model Hamiltonian with length = 9 base-pairs. **(b)** Transmission extracted at the midgap and plotted as a function of the DNA length. Calculation parameters are $\Gamma_B$ = 100 meV and λ = [10,50,100,500] meV, and $\Gamma_{L(R)}$ = 600 meV for all cases.

## 4. The Impact of the Decoherence Probe Self-Energy Treatment

In this section, we look at the density of states of the *model Hamiltonian* to understand the physical implication of the real part of the self-energy. The real and the imaginary parts of the self-energy are related by the Kramers-Kronig relation (equations (13)-(16)). The mathematical form of the self-energy used is dictated by the physical need to decrease the unphysically large DOS in the bandgap of the DNA induced by the E-indep model. The real part of the self-energy should be calculated from the imaginary part, and this will require the evaluation of a large number of integrals.

To understand the physical implication of including $Re[\Sigma_B(E)]$, we first plot the real and imaginary parts of the self-energy in Figure 4 for an onsite potential $\epsilon_m = 0$ eV. The real part of the self-energy $Re[\Sigma_B(E)]$ shifts the onsite potential $\epsilon_m$. The imaginary part is directly proportional to the scattering rate due to the decoherence probe, which causes a broadening of the density of states. Notice how the amount of shift and broadening decays at energies away from resonance ($\epsilon_m = 0$ eV). Next, we quantitively compare the density of states of the model Hamiltonian: (i) using the real and imaginary parts of the decoherence self-energy as defined in equation (13), and (ii) omitting the real part and defining $\Sigma_B(E) = i \times Im[\Sigma_B(E)]$.

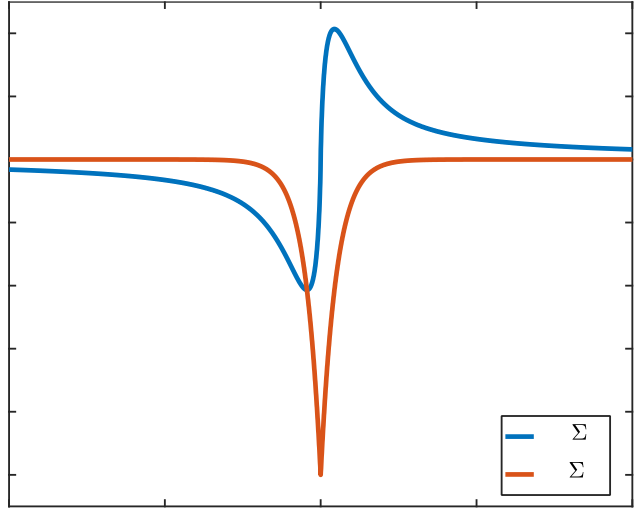

Figure 4 The real and imaginary parts of the self-energy obtained from evaluating equations (13)-(16), with $\Gamma_B = \lambda = 100$ meV.

For this example, we consider the model Hamiltonian with 9 bps (the 3'-$C_3GCGC_3$-5'). In Figure 5, we can see that the impact of including the real part of the self-energy is to shift the DOS peaks by approximately the value of the $Re[\Sigma_B(E)]$. However, as we enter the HOMO-LUMO gap, the impact of the real part is minimal and both treatments yield similar results. Similarly, the impact on the resulting transmission follows the same trend as the DOS (see Figure 17). We then calculate the integral of DOS from -2000 eV to 2000 eV, with a fine mesh of 1 meV around the onsite potentials [$\epsilon_{H(L)} \pm 1.5$ eV], and a coarse mesh of 10 meV otherwise. The total number of energy levels is 14 (two per block). Thus, the expected integral of DOS is 14 (excluding the electron spin multiplicity). Including both the real and imaginary parts of the self-energy yielded $\int DOS\ dE = 14$, as expected. However, including only the imaginary part of the self-energy yielded $\int DOS\, dE = 13.461$. This result implies that the total density of states is not conserved when scattering is included, which is incorrect [53].

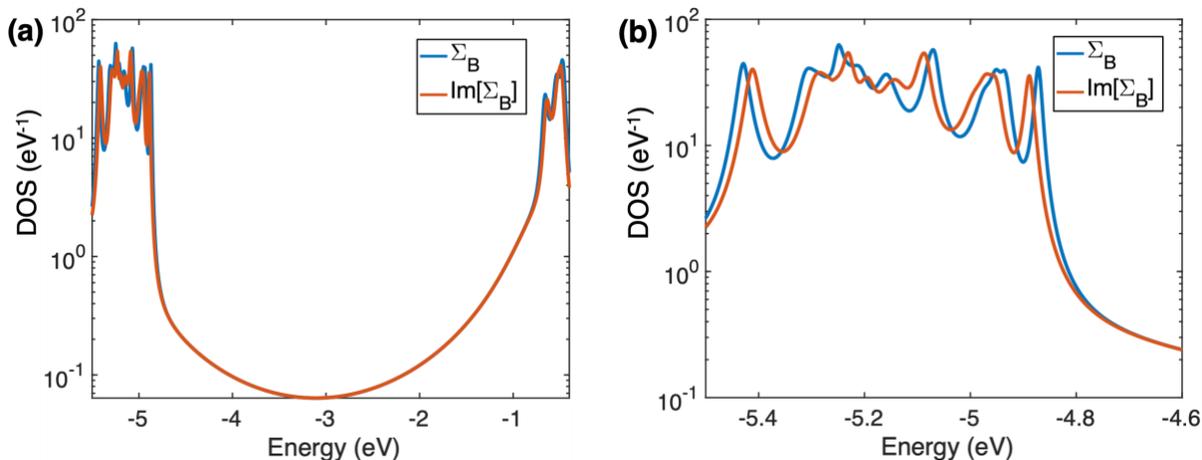

Figure 5 Density of states plots for the model Hamiltonian of 3'-C$_3$GCGC$_3$-5' using the full complex decoherence probes self-energy and only the imaginary part of the self-energy. **(a)** The DOS plot over the energy spectrum displaying HOMO, HOMO-LUMO gap, and LUMO regions. **(b)** Zoom-in to the edge of HOMO region. The calculation parameters are $\Gamma_B = 100$ meV, $\lambda = 100$ meV, and $\Gamma_{L(R)} = 600$ meV.

Although the full self-energy treatment is essential to yielding the correct integral of the DOS, the qualitative and quantitative differences in energy windows around the HOMO are small. In the full treatment of the system Hamiltonian obtained from DFT, the number of molecular orbitals per nucleotide is approximately 380. As a result, there are ~11,400 orbitals for a 15 base-pairs long strand that needs to be accounted for when using Kramers-Kronig relation (equations (13)-(16)) in the full self-energy treatment. Further, the results of this section show that we can safely neglect the real part of the self-energy for calculations that do not require a self-consistent integration of DOS to achieve convergence. Therefore, in the remainder of this article, we neglect the real part and use $\Sigma_B(E) = i \times Im[\Sigma_B(E)]$ to model the *energy-dependent* decoherence.

## 5. Decoherence and Temperature

In this section, we discuss how the E-dep decoherence model can reflect the effect of temperature in its current treatment. The usage of the λ parameter has been explored before in solid-state semiconductors, where λ is treated as the Urbach energy[54]. In terms of the Urbach energy, λ was mainly used to account for DOS broadening at the bandgap tail of the semiconductors. Previous work has also shown that λ is directly proportional to temperature [55,56]. Therefore, as temperature increases, we expect $\lambda$ to increase, which agrees with intuition. Additionally, DOS broadening is directly proportional to the maximum decoherence rate ($\Gamma_B$). Therefore, we also expect $\Gamma_B$ to increase with temperature. However, the precise mathematical forms of $\lambda$ and $\Gamma_B$ would depend on the Hamiltonians governing the scattering mechanisms (fluctuations in hydration shell, solvent-ions movement, vibronic coupling fluctuations, …etc.), each of which is a challenging problem worthy of its own investigation.

We plot the self-energy of the *E-dep* model using equations (14)-(15) at different $\Gamma_B$ and $\lambda$ values in Figure 6. As discussed above, we expect both values to increase with increase in temperature. Increasing $\Gamma_B$ increases the peak value of both the real and imaginary parts of the self-energy, which corresponds to the maximum energy level shift and decoherence rate, respectively. Increasing $\lambda$ decreases the decay rate of the self-energy off-resonance. Comparing this plot with the results obtained previously by Gutierrez et al.,[24] yields interesting observations. Ref [24] uses the temperature-dependent phonon bath approach in modeling charge transport through DNA. Specifically, we look at Figure 4 from [24] that shows the temperature dependence of the real and imaginary parts of the Polaronic Green's function (referred to as **P**(E) in [24]). The two models have a similar trend: the effect of increasing the temperature on **P**(E) is proportional to increasing $\lambda$ in our *E-dep* decoherence model. One key difference, however, is seen in the peak values of the self-energies. In Ref [24], the temperature increase lowers the peak of **P**(E), whereas we expect it to increase the peak of $\Sigma_B(E)$.

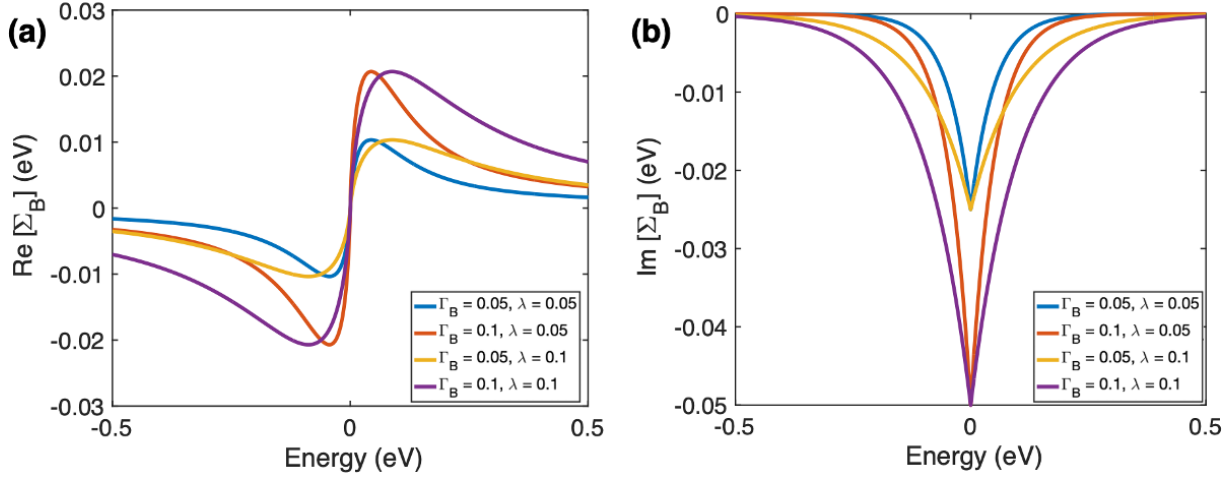

Figure 6 The real **(a)** and imaginary **(b)** parts of the self-energy for the E-dep decoherence model, which obeys the Kramers-Kronig relationship. The onsite potential ($\epsilon$) is assumed to be 0 eV, the values for $\lambda$ and $\Gamma_B$ are in eV.

To address this observation, we note that there are some fundamental differences in the systems studied. Gutiérrez et al. used a dissipative phonon-bath technique where the DNA Hamiltonian is represented by single energy levels for the base and the backbone, respectively. In addition, the phonon bath is only coupled to the energy level of the backbone. Our approach uses the full DNA Hamiltonian, and we apply the decoherence probes to all atoms. The reason for this is that DNA is a floppy molecule that exists in a solvent environment, usually consisting of at least some water and salt counterions. Therefore, decoherence in the DNA can occur due to the movement of water molecules and surrounding ions, and the vibrations of the DNA nucleotides and backbone. Thus, in contrast to Ref [24], we apply the decoherence probes to all atoms. We note that this is also an approximation because the precise form will be determined by the details of the Hamiltonian governing the scattering mechanisms, an unknown at this time. In addition, to understand how the decoherence self-energy would behave at different temperatures, let us consider the following temperature limits for the DNA system: 1) If the temperature is below 273 K, the water molecules freeze, and the effect of the fluctuating environment decreases. 2) If the temperature is more than 333 K, the DNA melts and loses its double-strand structure [57], and therefore the overlap between the orbitals ($\pi$-$\pi$ stacking) diminishes and the conductance decreases. 3) If the system is at room temperature (i.e., between the two limits), we found through our previous work that decoherence helps increase the conductance, reaching conductance values that agree with experiments [22,23]. The broadening of energy levels due to decoherence increases the conductance off-resonance. The increase in broadening is directly proportional to increasing $\Gamma_B$ and $\lambda$ in our *E-dep* decoherence model, which should also be directly proportional to the increase in temperature within the limits (273 K < T < 333 K). This conclusion agrees with other related work on 1D-disordered systems [58,59].

Another observation from the comparison is that our model of decoherence self-energy yields a Lorentzian shape for the imaginary part as opposed to the Gaussian shape seen in [24]. The difference of choice in using either of the two functions in modeling is not new in literature. Prior work in modeling decoherence in solid-state semiconductors has used Lorentzian functions [60,61]. More broadly, previous experimental work in extracting the DOS of organic semiconductors using XPS measurements has shown curve-fitted data to Gaussian, Lorentzian, and even a mixture of both functions [62,63]. In terms of modeling decoherence in DNA or other weakly-coupled systems, studying the effect of changing the decoherence shape from Lorentzian to Gaussian or to other functions are useful directions for future work.

## 6. The DNA System

The molecular building blocks of π-stacked and conjugated organic semiconductors with fixed atomic coordinates have sharp energy levels. In bulk form, these molecular building blocks are weakly coupled. It is experimentally known that these semiconductors have a broadened density of states with an exponentially decaying tail into their bandgap [62–66]. Several factors contribute to the broadening: thermal disorder (electron-phonon interactions), structural disorder, dopant impurities, and randomly distributed defects.

A single DNA strand consists of π-stacked aromatic rings and has a bandgap of 3-4 eV. The distribution of energy levels is sequence-dependent and varies along the length of the strand (see Figure 7) The non-rigid nature of the DNA molecule and its ionic/solvent environment results in a broadening of its molecular orbitals. Previous work showed that capturing the effect of this energy level broadening is necessary to explain experiments [23]. The *E-indep* decoherence model used in the previous work missed important aspects (the unphysical increase of transmission in the bandgap) that we discussed in section 4.

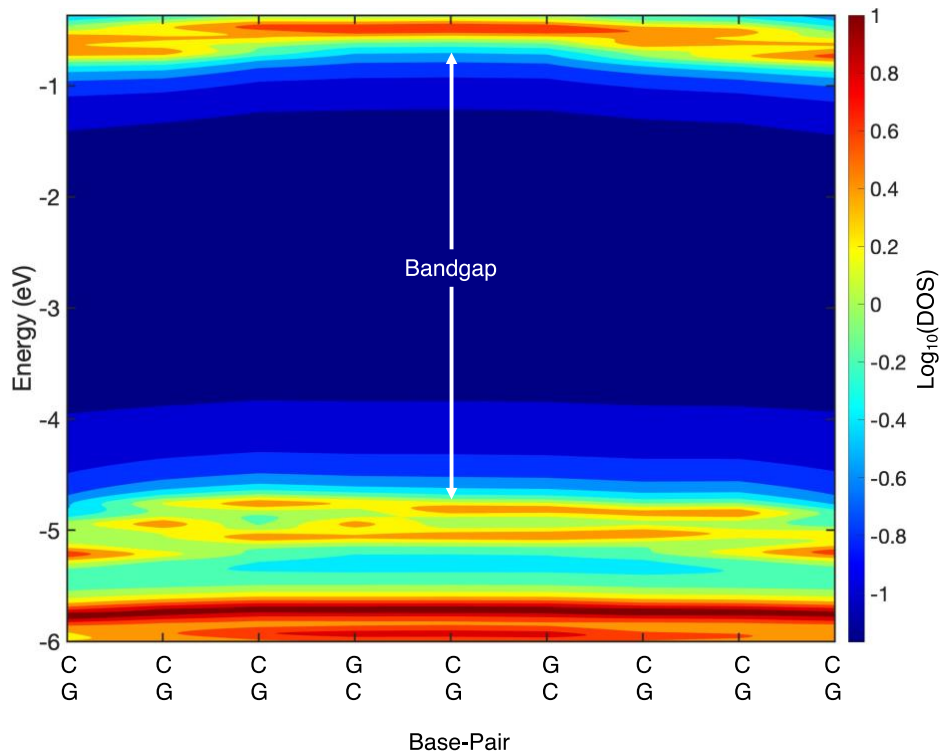

Figure 7 2D DOS plot of 3'-C$_3$GCGC$_3$-5'. The plot shows the nonuniform spatial and energy distribution of the molecular orbitals along the length of the strand. The calculation parameters are $\Gamma_B$ = 100 meV, $\lambda$ = 50 meV, and $\Gamma_{L(R)}$ = 600 meV.

In applying the E-dep model to DNA, a challenge is determining the values of $\Gamma_B$ and $\lambda$. Therefore, we look at the literature to determine the decoherence rates of other organic molecules. Ref [63] found the DOS broadening of aromatic molecules such as pentacene reach 300 meV. In addition, Parson's work [67,68] estimates the decoherence rates for solvated aromatic molecules to be in the 60-130 meV range. Using similar decoherence values in the *E-indep* model heavily broadens the DOS and washes out the energy dependence of transmission in DNA and makes the transmission in the bandgap unphysically large. That is why it is crucial to consider the *E-dep* model.

### A. System Under Study

In this section, we study the ds B-DNA sequence 3'-C$_3$G[CG]$_l$C$_3$-5', with $l$ = 1-4. We modeled four B-DNA strands using the full Hamiltonian resulting from density functional theory (see section 3).

The strands are 9-mer, 11-mer, 13-mer, and 15-mer, which correspond to varying $l$ = 1-4. For the charge transport calculations, we set the contacts to be at the cytosine nucleotides at both the 3'- and 5'- ends of the strands (as shown in Figure 1a). This setting is based on the previous experimental work [34], where they attached the thiol-linker groups that connect the gold contact tips to the DNA at the 3'- and 5'- end cytosines. We vary $\Gamma_{L(R)}$ from 100 meV to 10,000 meV and find that our results do not vary significantly for $\Gamma_{L(R)}$ > 600 meV (see Figure 19). We are interested in a physical scenario where the contact resistance is smaller than the intrinsic resistance of the DNA. Hence, we set the contact coupling to be 600 meV in the following calculations.

### B. DOS and Transmission

To highlight the main differences between the *E-dep* and *E-indep* models, we start with examining the density of states of $l$ =1 case and vary the exponential decay term λ = [50, 100, 150, and 200 meV] while keeping the decoherence parameter fixed at $\Gamma_B$ = 100 meV. We plot the DOS in the HOMO region near the HOMO-LUMO gap in Figure 8, and it shows that the *E-indep* model heavily broadens and washes out the variation between high (peaks) and low (valleys) DOS regions. We also see that for the *E-dep* model, as λ increases, the DOS peak broadening increases. As a result, the nearby peaks start to merge. The low-DOS regions (valleys) deep inside the occupied energy levels (E~-5.5 eV) and in the bandgap (E~-4.5 eV) are higher than their *E-dep* counterparts by more than an order of magnitude. The transmission results are consistent with these observations. Figure 9 shows the transmission plot when $\Gamma_B$ = 100 meV while changing λ. We notice that the *E-dep* model has relatively distinct peaks and valleys.

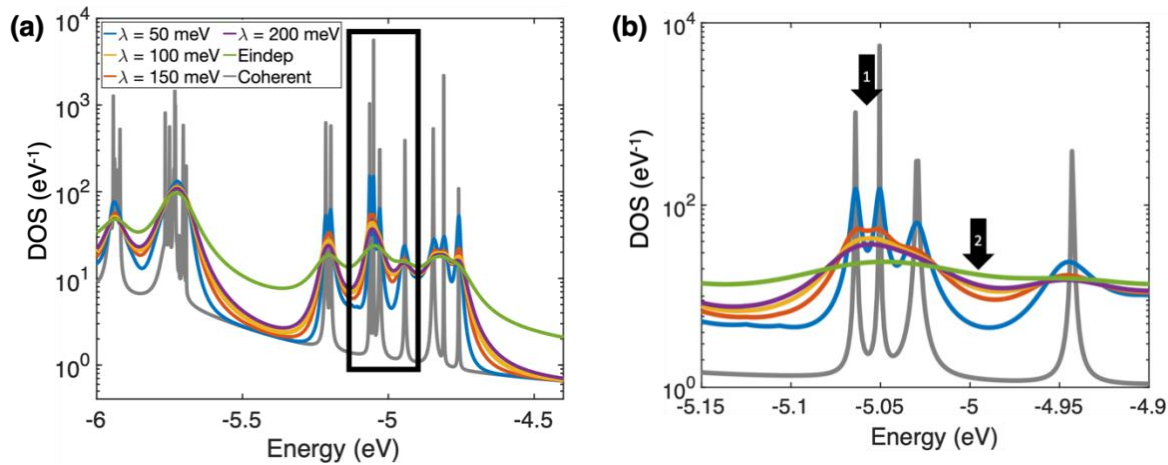

Figure 8 (**a**) Total DOS for $l$=1 (3'-C$_3$GCGC$_3$-5') in the HOMO region (E < -4.759 eV) near the HOMO-LUMO gap. (**b**) The inset shows the highlighted region with arrows pointing at the two main observations: 1) increasing $\lambda$ increases the broadening and nearby peaks start to merge, 2) Increasing $\lambda$ increases low-DOS regions found between the high-DOS peaks due to the higher broadening.

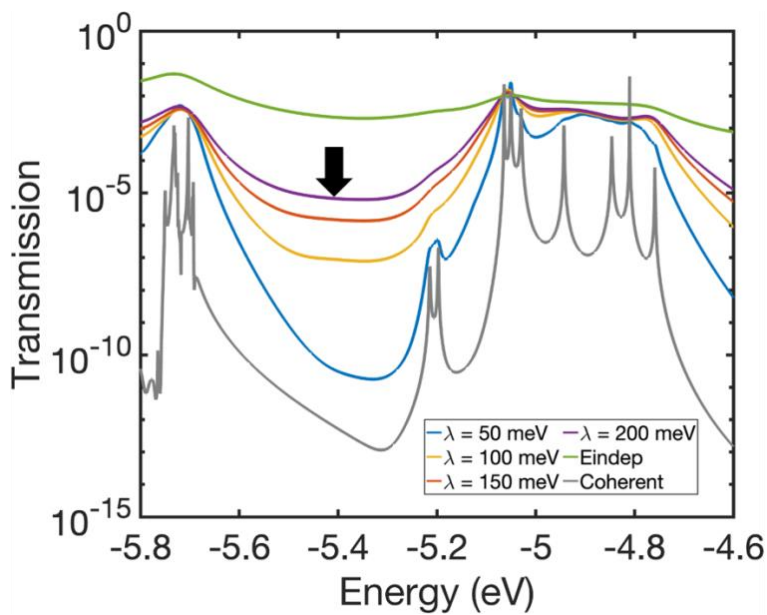

Figure 9 Transmission vs energy for 3'-C$_3$GCGC$_3$-5'. Solid grey line corresponds to the coherent transport, for both *E-dep* and *E-indep* models $\Gamma_B$ = 100 meV. The arrow is pointing at the transmission in an energy gap between two mini-bands (~-5.7 eV and -5.2 eV) within the HOMO region.

We plot the DOS for a molecular orbital localized at the first three CG base-pairs at the 3'-end of the strand in Figure 10a. The calculated full-width-half-maximum (FWHM) of the DOS peak as a function of $\lambda$ is shown in Figure 10b. We find that the peak width increases monotonically with increasing $\lambda$ and approaches the line shape of the *E-indep* model.

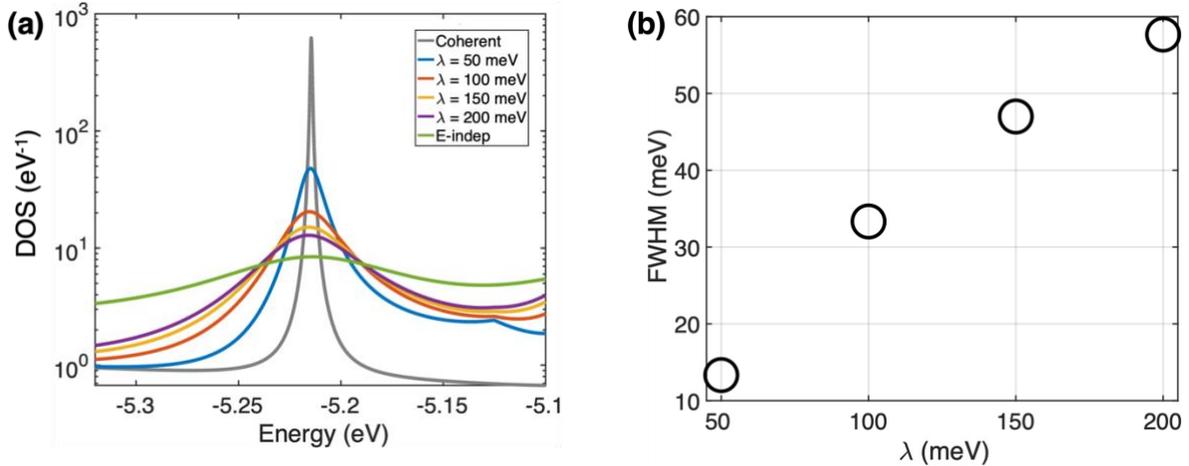

Figure 10 DOS of molecular orbital localized at the first three GC base-pairs within the strand 3'-C$_3$GCGC$_3$-5'. (a) DOS of Molecular orbital at -5.225 eV. (b) Full-Width-Half-Maximum of the DOS peak under different $\lambda$ values, with $\Gamma_B$ = 100 meV.

In our DNA calculations, we find that if $\lambda$ is set to 10 meV or less, the transmission spectrum would be closer to the coherent transport limit (see Figure 18 in Appendix C). In addition, as we have discussed above, significantly increasing $\lambda$ can cause the transmission near the midgap to become unphysically large, acting more like the *E-indep* model. In addition, as discussed in section 6, when temperature increases, we expect $\lambda$ to increase. In our modeling of DNA charge transport, we assume room temperature as in most conductance measurement setups. Therefore, in the following section, we limit $\lambda$ to be within 10-150 meV.

### C. DNA Conductance versus Length

As a case study for the *E-dep* model, we investigate the relationship between the DNA strand length and the conductance. We consider four B-DNA strands 3'-C$_3$G[CG]$_l$C$_3$-5' ($l$ = 1-4) and study the conductance as a function of length. Experimentally, this family of strands shows an exponential decay in conductance with length, but modeling yielded a much weaker dependence of conductance with length [34]. A likely reason for this inadequacy is the large density of states in the bandgap as a result of using the *E-indep* decoherence model. We showed in Figure 3 that the *E-indep* model indeed leads to an incorrect dependence of transmission versus length.

We model the strands by fixing the decoherence rate at $\Gamma_B$ =100 meV and let $\lambda$ range from 10 to 150 meV. We calculate the conductance using equation 26. We assume that 1) the comparison between the gold work function (5.3 eV) and the ionization potential of the DNA nucleotides puts the expected contact

Fermi energy ($E_f$) closer to the HOMO in the HOMO-LUMO gap [22], and 2) we are operating at the low-bias regime; hence, it is unlikely that $E_f$ will go below the HOMO (i.e., $E_f \geq$ HOMO). The Fermi energy is difficult to determine because it depends on the details of the contact geometry. As a result, we are guided by the approach in Ref [69], which involves a) sweeping the Fermi energy to gain an understanding of the experimental system being modeled and b) determining the Fermi energy when the calculated conductance is equal to the experimental value (see Figure 11). The results display low sensitivity to λ except at 10 meV, where for $l$ = 1 and 2, the conductance is lower than the experimental value over the entire HOMO-LUMO gap (Figure 11 a,b). From this analysis, we plot the extracted conductance vs. strand length as shown in Figure 12a. The length is calculated using the relation 3.4 Å × number of base-pairs, where 3.4 Å is the axial rise in B-DNA (the shortest distance between neighboring bases along the helical axis of a strand). The motivation is to compare with the experimentally reported exponential decay of conductance with strand length, having a decay factor (β) of 0.20 Å$^{-1}$ [34].

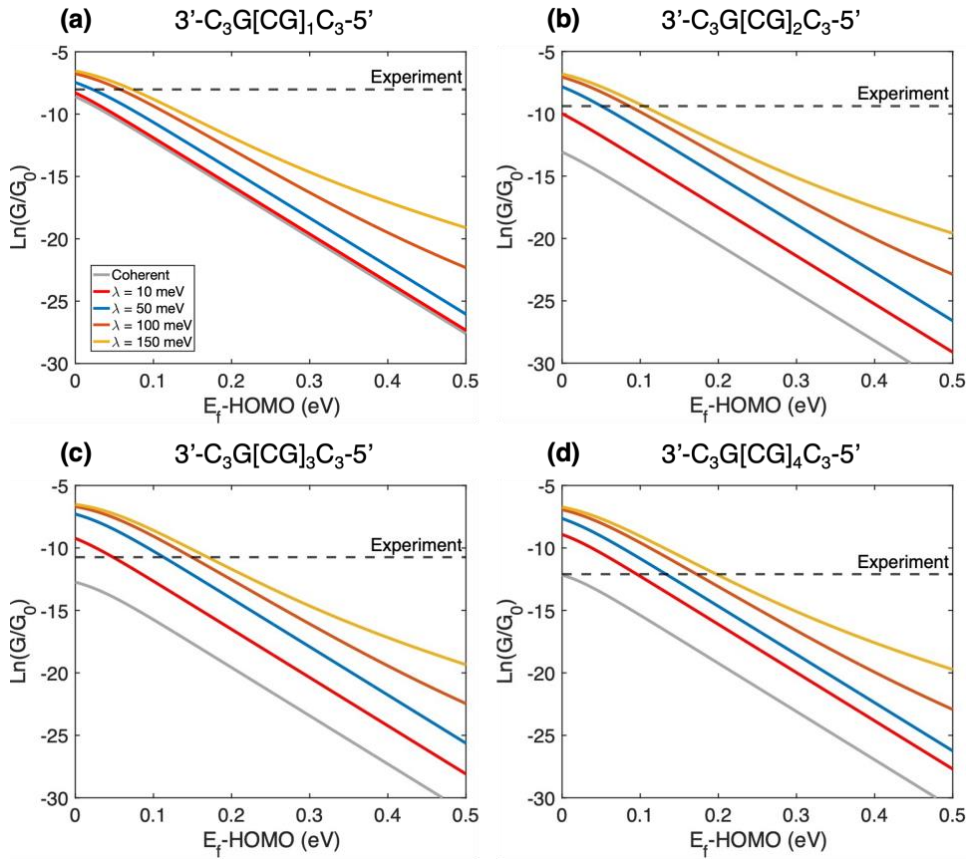

Figure 11 Conductance of the four strands ($l$=1-4) at different $\lambda$ values with the coherent transport for comparison. The dashed line corresponds to the targeted experimental conductance value. The trend shows $\lambda$ impact on the conductance decay rate as we enter the HOMO-LUMO gap.

Our calculations further reveal that the modeled value of conductance matches experiments at higher values for the Fermi energies as $\lambda$ increases (Figure 12b). This trend can be explained by Figure 11, which shows the tail of the conductance increasing in amplitude with the increase in $\lambda$ in the HOMO-LUMO gap. We have shown through equation 16 and Figure 3 that $\lambda$ dictates the exponential decay into the bandgap. Therefore, the conductance at the edge of the bandgap increases with increasing $\lambda$.

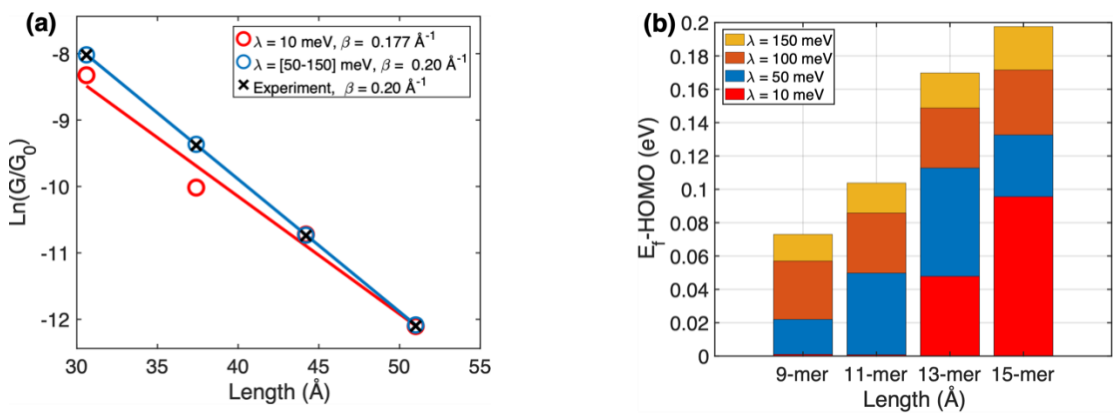

Figure 12 Conductance as a function of strand length. **(a)** Conductance trend at $\Gamma_B$ = 100 meV, $\lambda$ = [10,50,100,150] meV compared with experiment. The lines correspond to the curve fitting of the data points. The blue curve represents the results for $\lambda > 10$ meV. **(b)** $E_f$ of the extracted conductance values with respect to HOMO of the strand.

The second trend we observe from Figure 12b is that as the strand length increases, the Fermi energy at which the modeled conductance matches experiments moves farther away from the HOMO (for a fixed value of $\lambda$). To understand this, next, we discuss the effect of the contact-molecule junction on the expected Fermi energy location and how it is sensitive to the DNA length.

For a contact-DNA-contact system, partial charge transfer ($\delta$n) occurs from the contacts to the DNA due to the energy level broadening caused by the contacts and the misalignment between the Fermi energy of the contact ($E_f$) and the DNA Fermi energy ($E_{f_{DNA_0}}$) before the contact is established [69]. In turn, the molecular orbitals of the DNA shift to higher energies (the HOMO gets closer to the $E_f$ of the contacts). The relationship between the amount of shift in the molecular orbitals and $E_f$ due to partial charge transfer has been described in [69] as

$$E_f = E_{f_{DNA_0}} + \frac{\delta n}{n'} + U\delta n \qquad (27)$$

where $E_{f_{DNA_0}}$ is the Fermi energy of the DNA before forming the junction with the contacts, $\delta n$ is the amount of partial charge transferred, and $n'$ is the derivative of the electron number with respect to Fermi energy $(dn/dE_f)$, where $n = \int DOS(E)f(E - E_f)dE$. U is the charging energy, which is a material property that describes the change in the potential of the molecule per one added electron. In equation (27), the second term describes the impact of the density of states in the DNA on the Fermi energy shift (or the molecular orbitals shift). The last term describes the electrostatic interaction due to the charge transfer. To find the exact Fermi energy of the molecule after the contact-molecule-contact junction is formed, we require a self-consistent calculation that accounts for the contact details (atoms geometry and orientation), integral of DOS, and Poisson's equation. Although previous work shows the utilization of such a method on small molecules, the results yielded a wide range of Fermi energy locations extending from HOMO to LUMO [69]. This outcome is due to the sensitivity of the calculations to the DOS variation in the bandgap, the contacts geometry, and the choice of DFT exchange-correlation functionals. The difficulties faced in precisely calculating the Fermi energy of the molecule have shifted our focus into quantifying another critical parameter, the rate of the shift in energy per partial charge transfer.

We can utilize the second term in equation (27) to estimate the amount of molecular orbital shift per strand. we first extract the DOS from our calculations using equation (19). In addition, our results in section 5 show that ignoring the real part of the decoherence self-energy has minimal effect on the DOS in the HOMO-LUMO gap. Therefore, next, we integrate the DOS in the HOMO-LUMO gap as a function of Fermi energy,

$$n(E_i) = \int \sum_m DOS(m, E)\, f(E - E_i)\, dE \qquad (28)$$

where $m$ is the nucleotide number, and $f(E - E_i)$ is the probability of electron occupancy with respect to energy when $E_f = E_i$. Taking the inverse of the derivative $[dn(E_i)/dE_i]^{-1}$ will yield the rate of Fermi energy change as a function of electron number $dE_i/dn$ ($1/n'$ in equation (27)). Therefore, we can estimate the change in the DNA Fermi energy -which is a direct indicator of the amount of shift in molecular orbitals- as a function of electron number in the DNA.

This rate gives us another viewpoint of the expected trend in $E_f$ − HOMO. The higher the $dE_i/dn$ value, the higher the shift in the molecular orbitals, and the closer $E_f$ becomes to the HOMO of the DNA. The rate $dE_i/dn$ in the HOMO-LUMO gap is shown in Figure 13 and we listed the average values in Table 1. The results show that the expected molecular orbitals shift decreases as the strand length increases. This means that the longer the strand, the farther $E_f$ is expected to be from HOMO, which is consistent with the obtained trend seen in Figure 12b.

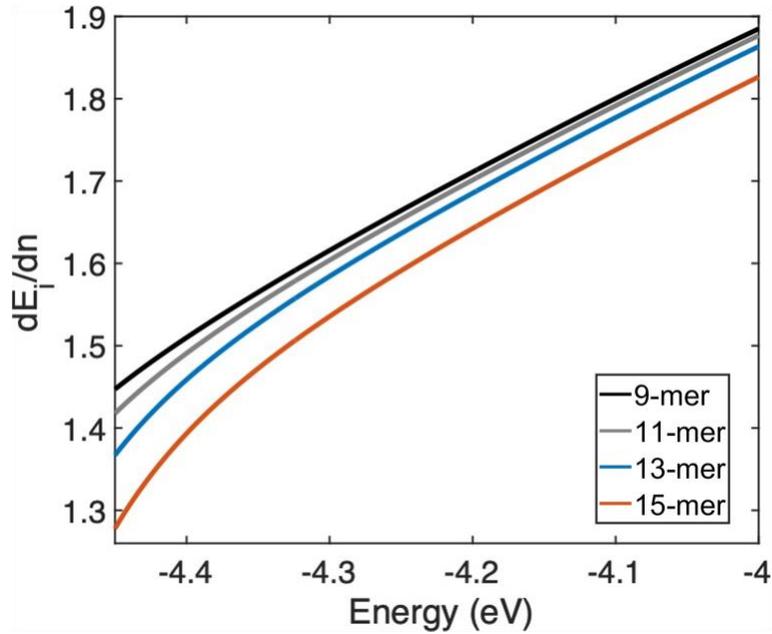

Figure 13 The resulting $dE_i/dn$ extracted from the HOMO-LUMO gap window for each strand, calculated at λ = 100 meV.

We also note that the trend is maintained for different λ values. Table 1 shows a comparison between λ = 50 and 100 meV. We notice that $dE_i/dn$ increases with decreasing λ. This can be understood as follows: as λ decreases, the energy levels broadening decreases, and the DOS decreases in the bandgap.

As a result, $n(E_i)$ decreases in the bandgap as well ($n$ is the integral of DOS) and its rate of change in energy ($dn/dE_i$) decreases. The reciprocal ($dE_i/dn$), however, increases.

Table 1 Comparison between the average rate of change in energy with respect to electron number ($dE_i/dN$) for 3'-C$_3$G[CG]$_l$C$_3$-5'($l$ = 1-4), $\lambda$ = 50 and 100 meV

| Strand | $dE_i/dn$ (eV/electron) $\lambda$ = 50 meV | $dE_i/dn$ (eV/electron) $\lambda$ = 100 meV |
| --- | --- | --- |
| (N=1) 9-mer | 1.9773 | 1.9750 |
| (N=2) 11-mer | 1.9697 | 1.9658 |
| (N=3) 13-mer | 1.9570 | 1.9503 |
| (N=4) 15-mer | 1.9211 | 1.9109 |

The low sensitivity on λ to yield conductance values within the experimental range provides us with an *energy-dependent* model that does not require over fitting. Of course, the ultimate goal for us is to derive a direct relation between the exponential decay term and the amount of broadening or decoherence in the system. However, this requires further work in both experiment and theory. From the given information in the literature about DOS broadening in molecular systems with a few hundred meV widths, the *E-dep* model is essential to apply such decoherence or broadening rates to yield physical results and maintain important transmission features. Further, DNA sequence affects the energy levels distributions along the strand. Therefore, we have applied the *energy-dependent* model on a sequence having 6 AT base-pairs and a single GC base-pair to show that the conclusions above are valid for a broader set of sequences. We refer the reader to the Appendix for more details. The results shown in this work further confirm the versatility of the new model and its applicability to simulate DNA strands.

## 7. Conclusion

Modeling decoherence for charge transport through DNA or other weakly-coupled systems has seen different approaches and levels of theory, such as: dissipative phonon-bath technique, vibronic dephasing, and solving time-dependent equations to account for temporal energy levels variation. However, these techniques were mainly applied to simplified Hamiltonians, consisting of single energy levels, due to their computational costs. Although these approaches can qualitatively describe DNA charge transport, using the full Hamiltonian that comprises a realistic DNA is essential to explain different aspects of DNA charge transport, such as: the effect of the backbone, methylation of nucleotides, single nucleotide mismatches, and intercalation.

One prominent model that can handle the full Hamiltonian (of hundreds of atoms) in a computationally efficient manner is the phenomenological decoherence probes method. In this study, we have introduced a phenomenological energy-dependent (*E-dep*) decoherent transport model that can be applied to weakly-coupled molecular structures such as DNA. The model overcomes the limitations of its previous energy-independent treatment as it allows for exponential decay of transmission with length at energies in the bandgap and maintains features in the transmission spectra within the valence and conduction bands. We applied the model on four B-DNA strands 3'-$C_3G[CG]_NC_3$-5'(N=1-4) and studied the conductance as a function of length. Previous experiments have shown that the conductance decays exponentially with the strand length. Using the *E-dep* decoherence model, we can model the experimental trends quantitatively without overfitting the decoherence parameters. The model can help study and design nanoelectronics devices that utilize weakly-coupled molecular structures. While a microscopic model for decoherence in DNA, a very challenging problem, is missing at this time, determining the Hamiltonians for the various microscopic processes identified will form a useful future study in researching realistic models for decoherence.

## 8. Acknowledgment

We acknowledge using the Hyak supercomputer system at the University of Washington. We also acknowledge NSF Semi Syn Bio Grant Number 2027165 and NSF Future of Manufacturing Grant Number 2229131.

## Appendix A: Choice of DFT Basis Set

In addition to the 6-31G(d,p) basis set, we have carried out calculations using 3-21G and a cc-pVDZ basis sets on the 9 base-pair strand (3'-C$_3$GCGC$_3$-5'). The calculations converge and we found the bandgap difference between cc-pVDZ and 6-31G(d,p) is 38 meV, and it is 10.6 meV between 3-21G and its 6-31G(d,p) counterpart. The results of the transport calculations are shown in Figure 14. The transmission curves are shifted to match the HOMO of each case. The calculation results show that the cc-pVDZ and 6-31G(d,p) basis sets have very similar trends (blue curve and black curve, respectively). On the other hand, we notice that the smaller basis set, 3-21G, shows the most variation in transmission in the HOMO region. Further, in studying charge transport through DNA, the ionization potential of the DNA bases or nucleotides is a figure of merit to compare between the quantum-mechanical calculation methods [39,40,70,71]. It is reported that B3LYP/6-31G(d,p) yields the correct trend as experiments, with an offset in values of about 300 meV [39]. Therefore, our choice of B3LYP/6-31G(d,p) is to achieve balance between accuracy and computation time.

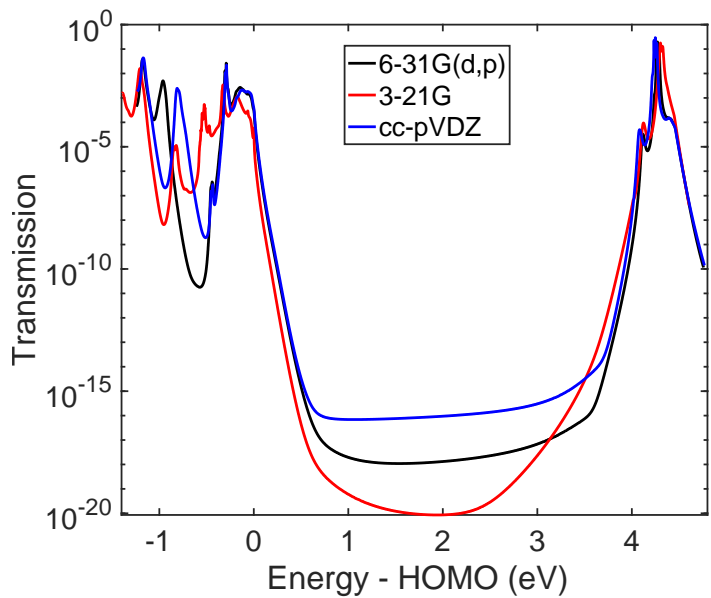

Figure 14 Transmission comparison of the different basis sets for the 3'-C3GCGC3-5. The x-axis is shifted to match the HOMO of each case. Calculation parameters: $\Gamma_L = \Gamma_R = 600 \; meV, \Gamma_B = 100 \; meV, \lambda = 50 \; meV$.

## Appendix B: Spurious Eigenstates in the DNA Bandgap

To apply the decoherence probes, we partitioned the system into nucleotides (backbone + base). In this procedure, we are setting the partition boundary at the covalent C-O bond between the sugar group at one nucleotide and the phosphate group of the neighboring nucleotide, respectively. To demonstrate how the backbone partitioning can induce spurious eigenstates in the bandgap, we tested two cases of neighboring cytosines: without a backbone ($CH_3$ terminated) (Figure 15a), and with a backbone (Figure 15b). We plot the molecular orbitals ($\epsilon_{k,m}$) of the Hamiltonian ($H$) after performing the partitioning with equation 7 in Figure 15c. The results show that the case with backbone has one eigenstate in the bandgap. Therefore, for this partitioning scheme, we expect to see a spurious eigenstate for every two neighboring nucleotides. For a ds-DNA with $N$ number of nucleotides, we have $N - 2$ spurious eigenstates appearing in the bandgap after partitioning.

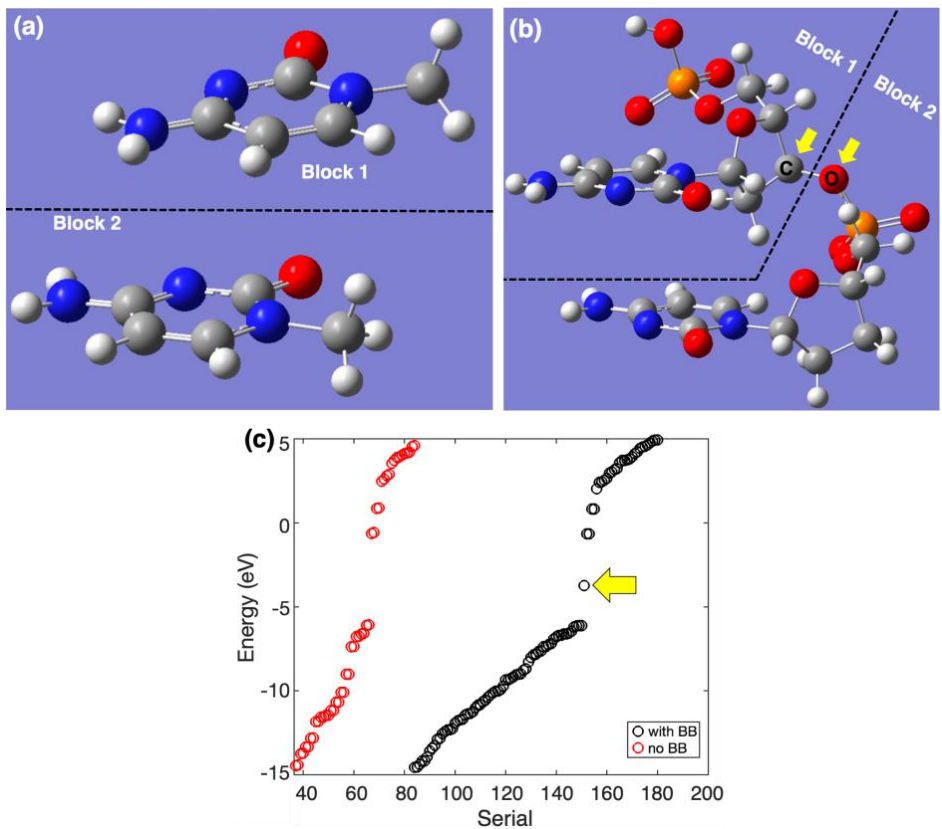

Figure 15 Two cytosine bases modeled **(a)** without the backbone (CH$_3$ terminated) **(b)** with the backbone. The arrows in **(b)** are poitning to the C-O bond at the boundary of the partitioning blocks. **(c)** The eigenstates of the partitioned Hamiltonians for the two cases. The arrow is pointing at the spurious eigenstate appearing the bandgap fro the case with the backbone.

For our modeled DNA strands, we find that the impact of the spurious eigenstates is to increase the transmission in the bandgap. The transmission plot in Figure 16 compares the effect of including and not including decoherence probes on the spurious states. The effect starts to appear ~600 meV away from HOMO, which is at least 400 meV higher than our determined $E_f$ location when fitted with experimental values (see Figure 10b in main text). In addition, the transmission amplitude at these spurious peaks is several orders of magnitude lower than the transmission in the HOMO region. Therefore, we did not apply the decoherence probes to them in our calculations.

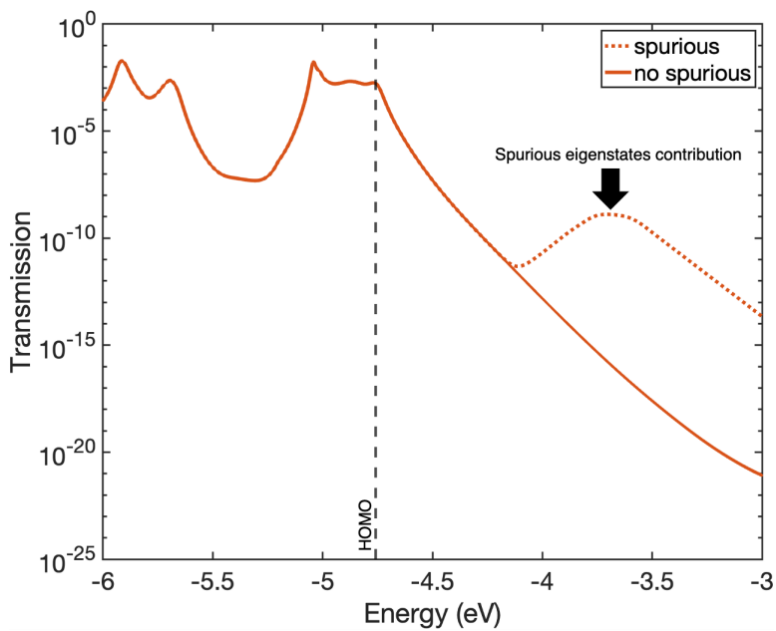

Figure 16 Transmission of 11-mer strand as an example for the effect of the spurious eigenstates on the transmission. It displays the comparison between including the spurious eigenstates into the decoherence probes versus not including them. The only impact on transmission starts to appear ~600 meV in the bandgap away from the HOMO. Calculation parameters are $\Gamma_B = 100$ meV, $\lambda = 100$ meV, and $\Gamma_{L(R)} = 600$ meV.

# Appendix C: Additional Transmission Plots

In this appendix we provide transmission plot for the model Hamiltonian discussed in section 4 as shown in Figure 17. In addition, we show a transmission plot of a full DNA Hamiltonian with a low λ value of 10 meV in Figure 18. For both plots, we have used the 3'-$C_3GCGC_3$-5' strand.

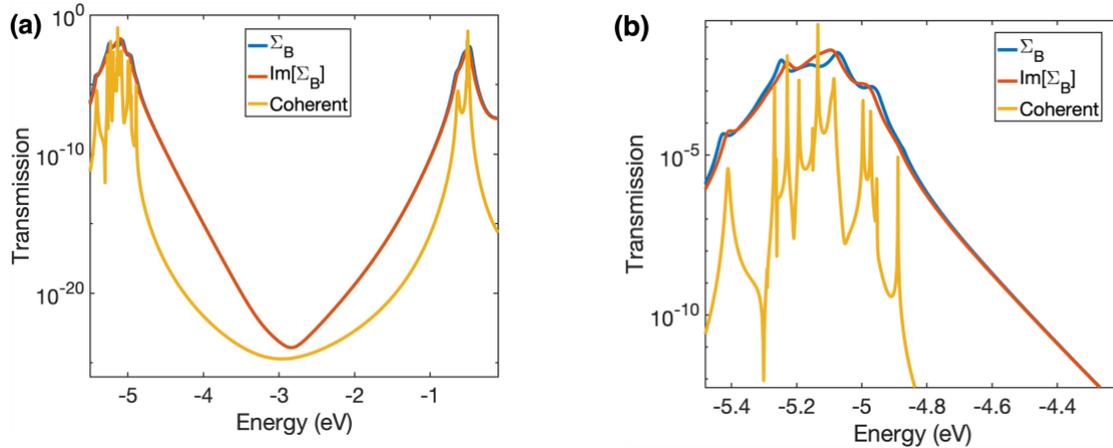

Figure 17 Transmission plot for the model Hamiltonian of 3'-$C_3GCGC_3$-5' using the full complex decoherence probes self-energy and only the imaginary part of the self-energy, in addition to the coherent transmission. **(a)** Transmission over the energy spectrum displaying the coherent transmission for comparison. **(b)** Zoom-in to the edge of HOMO region. The calculation parameters are $\Gamma_B$ = 100 meV, λ = 100 meV, and $\Gamma_{L(R)}$ = 600 meV.

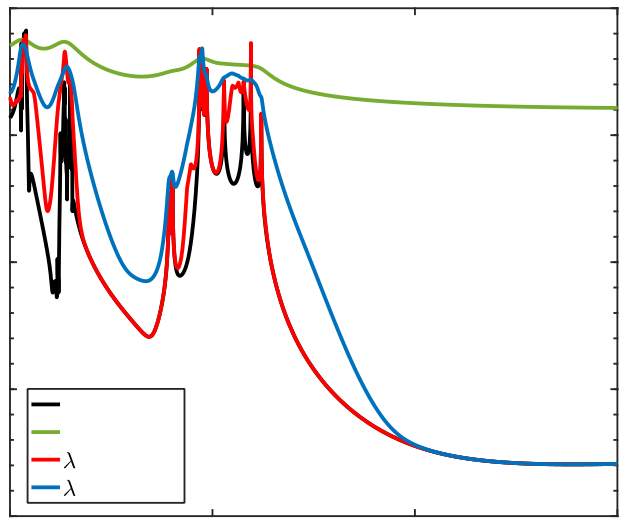

Figure 18 Transmission of 3'-C$_3$GCGC$_3$-5' strand comparing λ=10 meV with the coherent transport, λ=50 meV, and the *E-indep* model.

## Appendix D: Choosing the Contact Coupling

To minimize the impact of the contact coupling value on the results, we calculate the transmission for 3'-C$_3$GCGC$_3$-5' under different coupling rates $\Gamma_{L(R)}$ = [100-10,000] meV. The low value of 100 meV resembles weak coupling to the electrodes, whereas values between 500-1000 meV resemble strong coupling. We also added the extreme values of 5000 and 10,000 meV to see their impact on transmission. Figure 19 shows that the transmission increases with increasing the coupling value. We concentrated the analysis on the energy window extending from HOMO to HOMO+260 meV.

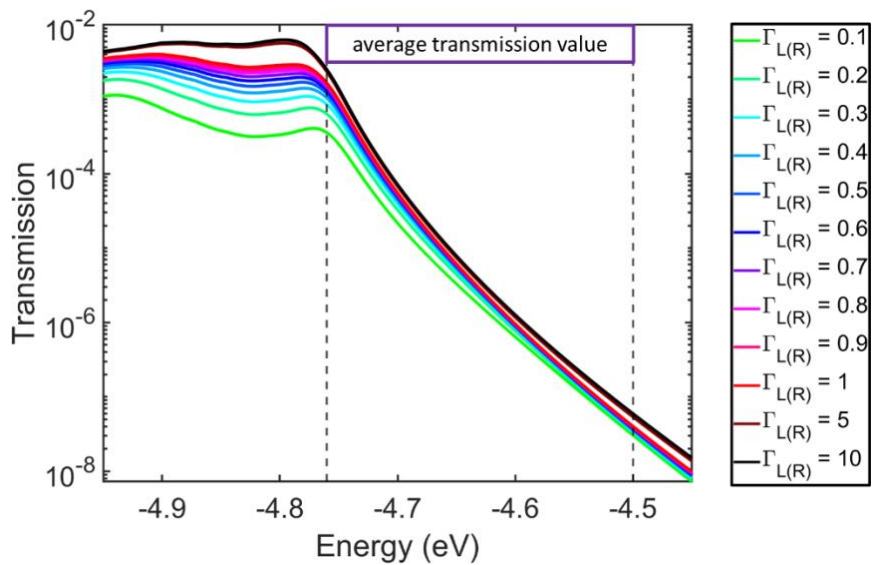

Figure 19 Transmission plot for 3'-C$_3$GCGC$_3$-5' at $\Gamma_B = \lambda$ =100 meV with different contacts coupling. The average transmission is calculated from the energy window HOMO to HOMO+260 meV. Legend shows the values in eV.

We first plot the transmission at HOMO as a function of $\Gamma_{L(R)}$ in Figure 20. We can see that increasing the coupling by 10 times from the low coupling limit of 100 meV to 1000 meV increases the average transmission by 4.5 times. We also observe that once we have a strong contact coupling ($\Gamma_{L(R)} >$ 500 meV), the increase in transmission starts to saturate. For instance, increasing the coupling from 500 meV to 1000 meV increases the transmission by less than 40%. Therefore, using $\Gamma_{L(R)} >$ 500 meV can be sufficient to limit the contact coupling impact on the transmission amplitude.

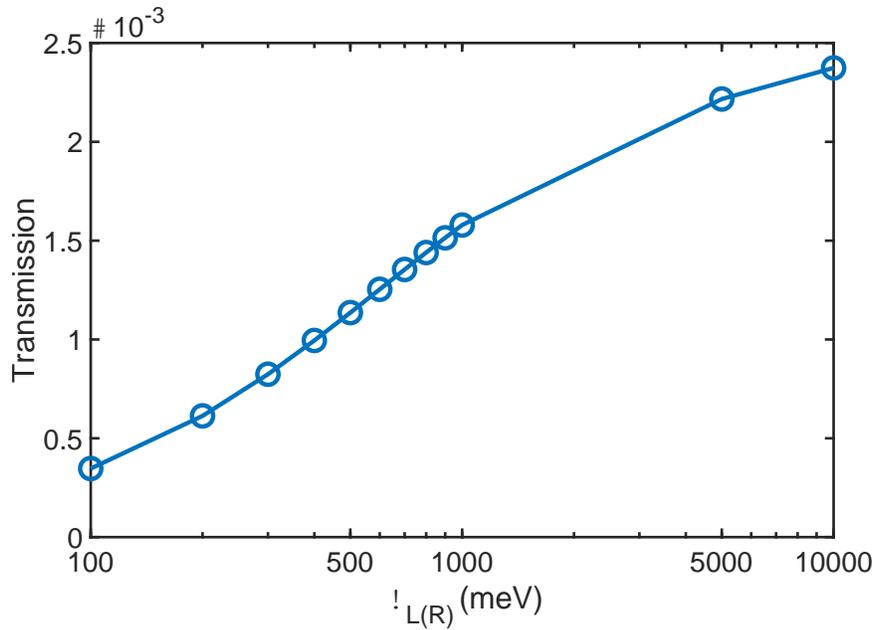

Figure 20 Transmission at HOMO as a function of contacts coupling. The x-axis is plotted in log scale for clarity.

Focusing on the coupling value of 600 meV, we plot the transmission increase with respect to $\Gamma_{L(R)} > 600$ meV (see Figure 21). The average increase in transmission (blue curve) is only 10% when we increase the coupling from 600 to 1000 meV (66.67%). In addition, the average increase in transmission is less than 50% when increasing the coupling to 10,000 meV (1600%).

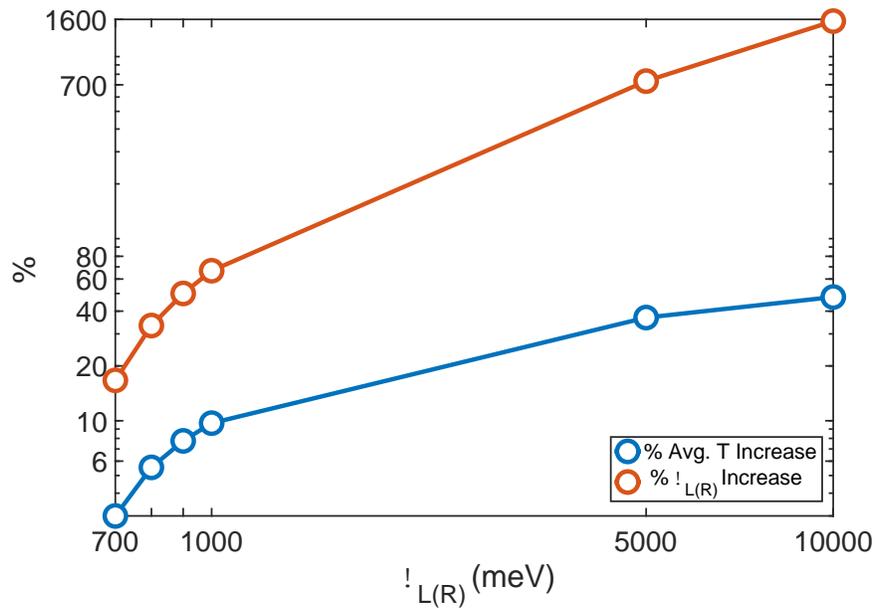

Figure 21 Percentage of Transmission (blue) and contact coupling (red) increase with respect to $\Gamma_{L(R)} = 600$ meV.

# Appendix E: E-dep Decoherence Model on an AT-rich DNA Sequence

To complement the GC-rich sequence in the main manuscript, in this appendix, we have applied the new *E-dep* decoherence model to the B-DNA 3'-TTTCTTT-5' sequence studied in[22]. This sequence has six AT base-pairs separated in the middle by a single GC base-pair. For simplicity, we will focus on the HOMO region this discussion. The six AT base pairs primarily determine the lower HOMO levels (the -5.9 eV to -5.8 eV miniband), while the single CG base-pair primarily determines the HOMO level (at -5.1 eV). The effect of the sequence on the energy levels distribution is reflected on the transmission profile seen in Figure 22, which shows a transmission increase in the miniband (-5.9 eV to -5.8 eV), followed by a decrease in transmission due to the lack of energy levels (low DOS region), then a sharp increase in transmission at -5.1 eV due to the HOMO of the GC base-pair. In general, the difference between the two models is that the *E-dep* decoherence model shows a more drastic transmission dip in low DOS regions (between the energy levels and in the bandgap) and a sharper increase at resonance. These findings are consistent with the strand considered in the main manuscript.

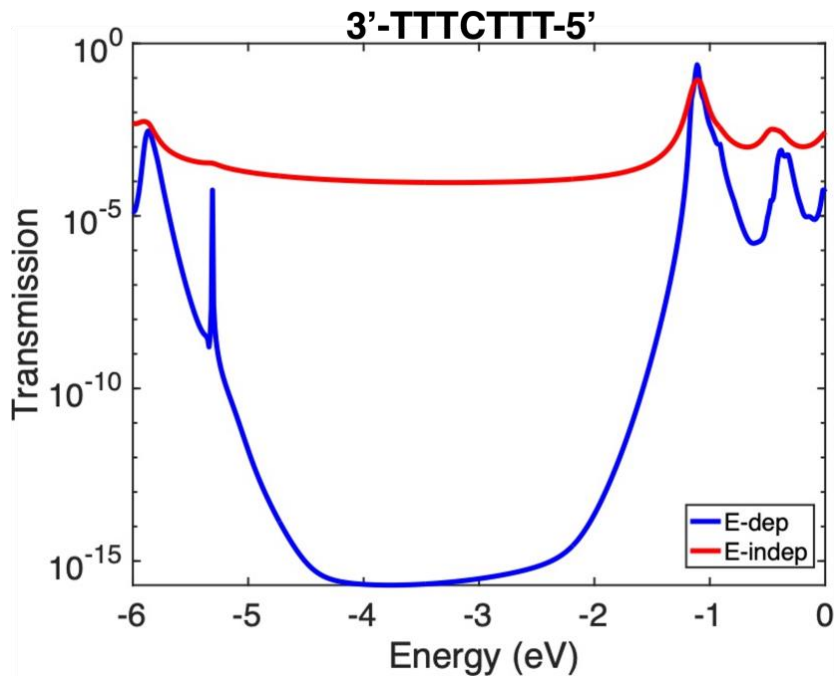

Figure 22 Transmission profile of the 3'-TTTCTTT-5'sequence spanning HOMO, HOMO-LUMO gap, and the LUMO regions. The decoherence parameters are $\varGamma_B = 100\ meV, \lambda = 100\ meV$.

## Appendix F: Divergence When Using NEGF To Simulate DNA

We attempted to model the DNA systems considered using the Non-Equilibrium Green's Function (NEGF) formalism, which can in principle provide physical insights into modeling decoherence using vibrational modes. This method is well-known to have the capability of explicitly including electron-phonon interactions. We tested this method using a DNA tight-binding *model Hamiltonian* generated from DFT calculations as described in section 3. Our main finding is that perturbation theory cannot be used to calculate the current through DNA because current conservation is violated in these weakly coupled systems where the decoherence strength is comparable to the coupling between bases in the Hamiltonian.

The $G^<$, $G^>$, $G^r$ and $G^a$ are calculated using the equations [72,73]:

$$[EI - H - \Sigma_L^r - \Sigma_R^r - \Sigma_{ph}^r]G^< = \Sigma^< G^a \tag{G1}$$

$$[EI - H - \Sigma_L^r - \Sigma_R^r - \Sigma_{ph}^r]G^> = \Sigma^> G^a \tag{G2}$$

$$[EI - H - \Sigma_L^r - \Sigma_R^r - \Sigma_{ph}^r]G^r = I \tag{G3}$$

$$G^a = [G^r]^\dagger \tag{G4}$$

where $\Sigma^{<(>)}$ is the self-energy matrix representing the electrons (holes) in-scattering (out-scattering). The self-energy $\Sigma^{<(>)}$ is composed of both contact leads and electron-phonon interactions. As for the acoustic phonon self-energy $\Sigma_{ph}^\alpha$, where ($\alpha = r, <, or >$), it is defined as

$$\Sigma_{ph}^\alpha(E) = D_a G_{q,q}^\alpha(E) \tag{G5}$$

where $D_a$ is the deformation potential and $G_{q,q}^\alpha$ is the diagonal element of the Green's function at layer q. Similar terms exist for phonons of finite energies [72,73]. Note that even these complex set of equations with diagonal self-energies is an approximation.

We considered deformation potential values of 10 and 100 meV. The calculations do not converge because DNA is a weakly-coupled system. That is, *electron-phonon interaction cannot be treated within the self-consistent Born approximation.* Other methods that go beyond the self-consistent Born approximation that work in weakly-coupled systems -like DNA- that do not violate current conservation are required. This requires further investigation that is outside the scope of this work as it will most likely require non-perturbative methods beyond toy models that can handle many orbitals at each DNA base.